# Can news and social media attention reduce the influence of problematic research?

Er-Te Zheng[1], Hui-Zhen Fu[2], Xiaorui Jiang[1], Zhichao Fang[3,4*], Mike Thelwall[1*]

* Corresponding author

Er-Te Zheng (ORCID: 0000-0001-8759-3643)
[1] Information School, The University of Sheffield, Sheffield, UK.
E-mail: ezheng1@sheffield.ac.uk

Hui-Zhen Fu (ORCID: 0000-0002-1534-9374)
[2] Department of Information Resources Management, Zhejiang University, Hangzhou, China.
E-mail: fuhuizhen@zju.edu.cn

Xiaorui Jiang (ORCID: 0000-0003-4255-5445)
[1] Information School, The University of Sheffield, Sheffield, UK.
E-mail: xiaorui.jiang@sheffield.ac.uk

Zhichao Fang (ORCID: 0000-0002-3802-2227)
[3] School of Information Resource Management, Renmin University of China, Beijing, China.
[4] Centre for Science and Technology Studies (CWTS), Leiden University, Leiden, The Netherlands.
E-mail: z.fang@cwts.leidenuniv.nl

Mike Thelwall (ORCID: 0000-0001-6065-205X)
[1] Information School, The University of Sheffield, Sheffield, UK.
E-mail: m.a.thelwall@sheffield.ac.uk

**Abstract**

News and social media are widely used to disseminate science, but do they also help raise awareness of problems in research? This study investigates whether high levels of news and social media attention might accelerate the retraction process and increase the visibility of retracted articles. To explore this, we analyzed 15,642 news mentions, 6,588 blog mentions, and 404,082 X mentions related to 15,461 retracted articles. Articles receiving high levels of news and X mentions were retracted more quickly than non-mentioned articles in the same broad field and with comparable publication years, author impact, and journal impact. However, this effect was not statistically signicant for articles with high levels of blog mentions. Notably, articles frequently mentioned in the news experienced a significant increase in annual citation rates after their retraction, possibly because media exposure enhances the visibility of retracted articles, making them more likely to be cited. These findings suggest that increased public scrutiny can improve the efficiency of scientific self-correction, although mitigating the influence of retracted articles remains a gradual process.

## 1. Introduction

Despite peer review and editorial oversight, problematic academic research continues to be published, including with issues due to misconduct, mistakes, or reliance on flawed prior research. Whilst automated tools have been developed to detect issues like text plagiarism (Wager, 2011) and image falsification (Bik et al., 2016), many problems require human intervention to identify, and some editors may need encouragement to engage in the often time-consuming process of initiating a retraction. The need for retraction may be flagged by the authors themselves or by other academics on post-publication review platforms like PubPeer. In some cases, flawed articles – such as those employing artificial intelligence to generate content (Zhang et al., 2024) or figures (Guo et al., 2024) – have been retracted following public discussion in the news or on social media. However, it remains unclear whether these instances represent exceptional cases or whether media and social media attention serve as a broader mechanism for post-publication correction of the scientific record.

Retraction serves as "a mechanism for correcting the literature and alerting readers to articles that contain such seriously flawed or erroneous content or data that their findings and conclusions cannot be relied upon" (COPE Council, 2019). Ideally, retractions should help reduce or halt the spread of misinformation originating from flawed articles. The extent to which retraction achieves this goal can be examined from two key perspectives: the *timeliness* and the *effectiveness* of retraction decisions. Retraction time lag – the duration between the publication of a flawed article and its subsequent retraction – is important, because delayed retractions allow misinformation to persist unchallenged for longer periods. Retraction effectiveness can be partly measured by post-retraction citations. Continued citations after an article has been retracted suggest that the retraction may not have fully curtailed its influence, assuming that these citations are not solely due to publication delays of citing articles or academic discussions explicitly acknowledging the retraction. Both the timeliness and effectiveness of retractions contribute to mitigating the negative impact of flawed articles (Zheng et al., 2024).

*1.1 Factors influencing retraction time lags and post-retraction citations*

Two factors associated with retraction delay lengths have been identified. Unsurprisingly, articles retracted due to misconduct have significantly longer retraction time lags (28.4-39.6 months) than those retracted for unintentional mistakes (22.7-24.0 months) (Nath et al., 2006; Steen, 2011). Whilst earlier studies found no correlation between journal impact factors and retraction time lags (Rai & Sabharwal, 2017), more recent research suggests that articles published in journals with higher impact factors have shorter retraction time lags (Madhugiri et al., 2021), perhaps due to such journals having greater resources to cope with retractions efficiently or stronger reputational incentives to address problematic publications promptly.

For post-retraction citations, previous research indicates that the later a paper is retracted, the more cited it tends to be (Sotudeh et al., 2022). Moreover, articles retracted due to plagiarism receive more post-retraction citations than those retracted due to falsification or fabrication (Dal-Ré & Ayuso, 2021). Additionally, retractions in journals with higher impact factors tend to attract more post-publication citations. For example, retracted articles from Q1 journals in the *Journal Citation Reports* initially have a rapid decline in citations following retraction, but this is often followed by a subsequent increase (Fu, 2016a, 2016b). Perhaps surprisingly, the number of retractions associated with an author does not seem to influence their post-retraction citations, however (Steen, 2012).

*1.2 News and social media attention towards retracted articles*

Whilst news and social media can both spread and counter misinformation (Akeriwe et al., 2023; Goel & Gupta, 2020), few studies have investigated their role in the discussion or dissemination of problematic articles (Khan et al., 2022), except in the context of COVID-19. This is an important gap, as news and social media coverage may influence not only the general public (Dempster et al., 2022; Ecker & Antonio, 2021) but also the scientific community. During the COVID-19 pandemic, news and social media coverage of subsequently retracted articles contributed to widespread misunderstandings about the virus amongst healthcare professionals, patients, and the broader public (Caceres et al., 2022). For instance, several articles about the side effects of COVID-19 vaccines were retracted due to unreliable data sources (Walach et al., 2021), but they continued to be cited by anti-vaccine advocates as evidence to oppose government mandates for nationwide vaccination (Brüssow, 2021; Prieto Curiel & González Ramírez, 2021).

A vast majority of retracted articles, along with their corresponding retraction notices, are discussed on social media after retraction (Serghiou et al., 2021). On platforms like X, retraction-related posts receive significantly more engagement – measured by likes, retweets, and comments – than comparable non-retracted articles (Dambanemuya et al., 2024; Peng et al., 2022; Serghiou et al., 2021). Notably, retracted articles tend to attract more attention than their official retraction notices (Serghiou et al., 2021). Highly cited retracted articles continue to attract considerable attention across social media platforms such as X, Facebook, and Mendeley, even after their retraction, highlighting their persistent influence (Jan & Zainab, 2018; Khan et al., 2022).

*1.3 Objevtives of this study*

Despite the brief summary of factors related to retraction time lags and post-retraction citations provided above, no studies appear to have specifically investigated the influence of news and social media discussions on retraction time lags and post-retraction citations. This study investigates whether frequent news and social media attention associates with a reduced influence of retracted articles. To address this question, we collected news and social media mentions for a set of retracted articles, divided into *highly mentioned* and *non-mentioned* groups, both before and after retraction. We then compared retraction time lags and post-retraction citation counts between these groups to identify any potential associations. The research questions (RQs) are as follows:

- RQ1: Do articles with greater news and social media attention before retraction get retracted more quickly?
- RQ2: Do the citation rates of articles with greater post-retraction news and social media attention decline more rapidly?

## 2. Data and methods

*2.1 Research workflow*

The overall research design was to compare differences in retraction speed and post-retraction academic impact between articles that received high versus no mentions in news and social media.

First, we collected pre- and post-retraction mention counts from news outlets, blogs, and X (formerly Twitter) for a set of retracted articles. Based on these mentions, the articles were categorized into two groups: *highly mentioned articles* (treatment group) and *non-mentioned articles* (control group). This classification was performed separately for pre- and post-retraciton mentions.

Second, to ensure comparability between the two groups, we employed the coarsened exact matching (CEM) method to select articles with similar characteristics from the highly and non-mentioned groups within both the pre- and post-retraction datasets.

Finally, we conducted two key analyses: (1) we compared the retraction time lag between highly and non-mentioned articles in the pre-retraction set; (2) we used a zero-inflated negative binomial regression model to analyze differences in pre- and post-citation changes between the two groups in the post-retraction set. These analyses allowed us to assess potential associations between news and social media attention and the effectiveness of retraction (Figure 1).

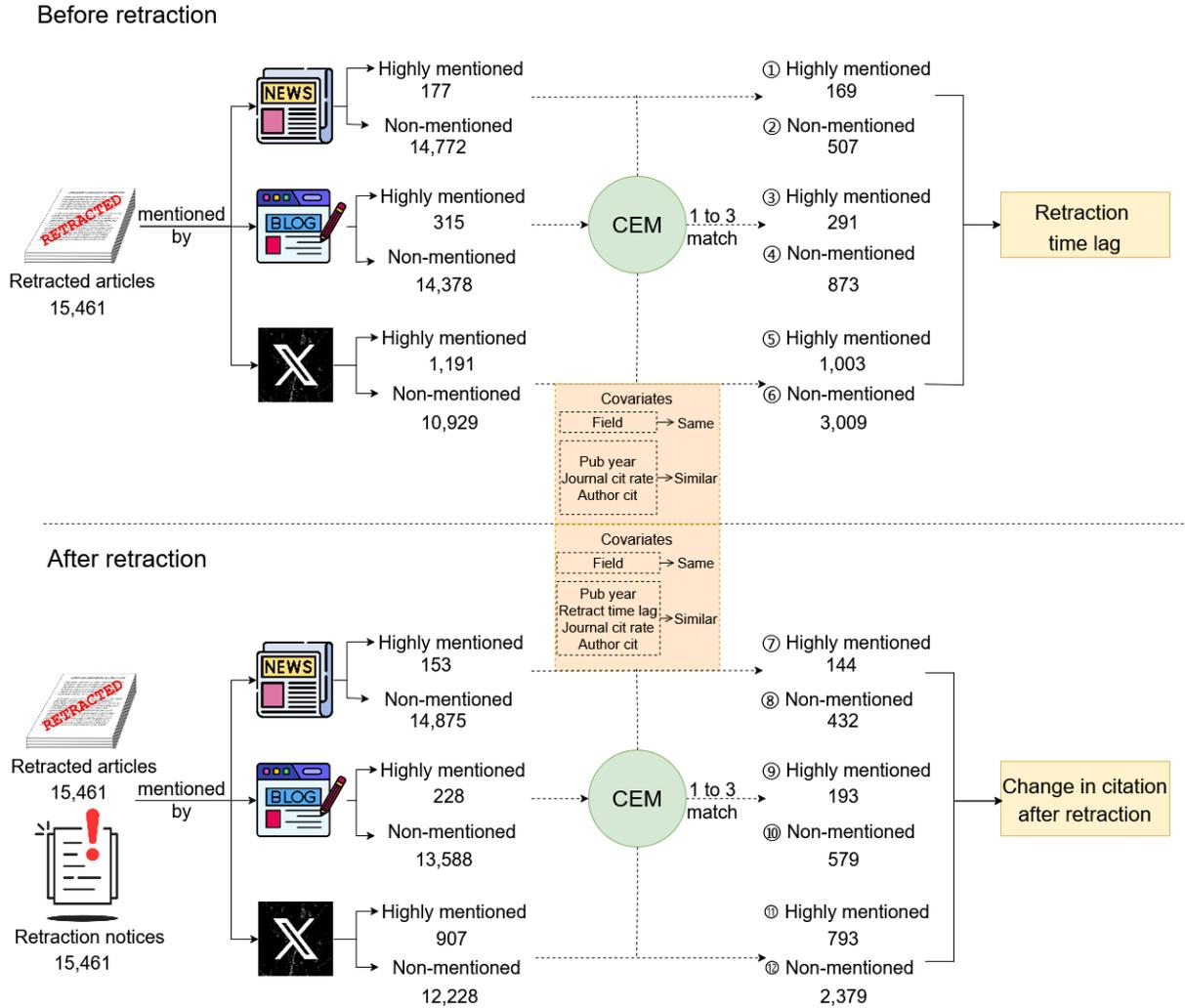

**Figure 1**. Overall research workflow. The study categorizes retracted articles into twelve groups based on their level of media mentions (high or none) across three media sources (news, blogs, and X) both before and after retraction.

## 2.2 Data collection

We used the scientific retraction-focused site *Retraction Watch* (https://retractiondatabase.org) to obtain a list of 15,461 retracted articles originally published between 2012 and 2021, along with their corresponding retraction notices. Retracted articles without retraction notices (2%) were excluded from the analysis. We also excluded bulk retractions, focusing exclusively on individually retracted articles. This decision was made because bulk retractions typically stem from a single underlying reason, rendering them non-independent events. Including such cases could introduce bias and compromise the accuracy of the analysis. Bulk retractions are instances where multiple scientific publications are retracted from a journal or conference simultaneously or within a short period. If a retraction notice includes two or more retracted articles, we classify it as a bulk retraction.

The majority of the retracted articles in our dataset are journal articles (Table 1), and falsification and manipulation is the most common retraction reason (Table 2).

**Table 1**. Document types of retracted articles.

| Article type | Number | Percentage |
|---|---|---|
| Journal article | 12,826 | 83.0% |
| Conference paper | 1,209 | 7.8% |
| Other | 1,426 | 9.2% |

**Table 2**. Retraction reasons of retracted articles. The categorization of retraction reasons follows Zheng & Fu (2024). Some articles were retracted for multiple reasons.

| Retraction reason | Number | Percentage |
|---|---|---|
| Falsification and manipulation | 3,908 | 25.3% |
| Mistakes | 3,287 | 21.3% |
| Self-plagiarism | 2,991 | 19.3% |
| Plagiarism | 1,327 | 8.6% |
| Authorship issues | 769 | 5.0% |
| Ethical issues | 585 | 3.8% |
| Reasons not available or uncategorizable | 4,423 | 28.6% |

We used DOI searches to collect the number of mentions of the retracted articles and their corresponding retraction notices in news outlets, blogs, and X, as recorded by Altmetric.com until November 2022. These three sources of altmetric attention are among the primary channels through which articles are likely to be disseminated (Fang et al., 2020; Trueger et al., 2015). It should be acknowledged that although Altmetric.com is one of the largest publicly available sources for tracking online attention (Karmakar et al., 2021; Meschede & Siebenlist, 2018), its coverage may be incomplete, likely biased towards English-language sources, and concentrated on platforms that support automated data extraction.

To quantify post-retraction attention, we collected mentions of both the retracted articles and their corresponding retraction notices. The combined number of mentions was used as a measure of the visibility of retractions (Haunschild & Bornmann, 2021). Although mentions of retraction notices differ conceptually from mentions of retracted articles, in practice virtually all post-retraction mentions of retraction notices allude directly or indirectly to the retracted articles. Therefore, combining the two provides a more comprehensive measure of post-retraction attention.

Metadata and citation counts for the retracted articles were gathered from the OpenAlex database via DOI searches conducted in August 2024. OpenAlex offers coverage comparable to that of the Web of Science and Scopus (Alperin et al., 2024; Culbert et al., 2024). The publication dates and retraction dates extracted from metadata were used to distinguish between pre- and post-retraciton phases for the retracted articles. Overall, the dataset comprised 10,671 news mentions, 2,674 blog mentions, and 288,357 X mentions before retraction; and 4,971 news mentions, 3,914 blog mentions, and 115,725 X mentions after retraction.

Although retracted articles generally attract considerable news and social media attention (Peng et al., 2022; Serghiou et al., 2021), there are still over half of them were not mentioned (Table 3), at least according to Altmetric.com, though Altmetric.com cannot track all news and social media mentions (Karmakar et al., 2021). Fewer than 10% of retracted articles were mentioned by news outlets before or after retraction, whereas more than 40% had at least one X mention before or after retraction.

**Table 3**. Number and percentage of retracted articles with different media mentions (2012-2021).

| Mention category | Mentioned by news (%) | Mentioned by blogs (%) | Mentioned by X (%) | Mentioned by one of them (%) |
| --- | --- | --- | --- | --- |
| Before retraction | 689 (4.5%) | 1,083 (7.0%) | 4,532 (29.3%) | 5,033 (32.6%) |
| After retraction | 587 (3.8%) | 1,873 (12.1%) | 3,233 (20.9%) | 4,219 (27.3%) |
| Before or after retraction | 1,013 (6.6%) | 2,484 (16.1%) | 6,186 (40.0%) | 7,125 (46.1%) |

*2.3 Matching control and treatment groups*

Retracted articles with news, blog, or X mentions exceeding a predefined threshold before retraction were classified as highly news/blog/X-mentioned articles, while those receiving no mentions in Altmetric.com were classified as non-mentioned articles. This classification process was repeated separately for articles after retraction, generating 12 sets of retracted articles (six pre-retraction and six post-retraction). Since media attention can change over time, there may be overlapping retracted articles between the two phases. For instance, an article categorized as non-mentioned before retraction could become highly mentioned after retraction, leading to potential overlap between the pre-and post-retraction sets. An article was classified as highly mentioned if its number of mentions on a given platform ranked within the top 30% of all retracted articles for that platform. Table 4 presents the specific threshold values used for classification.

**Table 4**. Number of highly mentioned articles and thresholds for high news, blog, and X mentions.

| Phase | Number of articles with high news mentions (threshold for high news mentions) | Number of articles with high blog mentions (threshold for high blog mentions) | Number of articles with high X mentions (threshold for high X mentions) |
| --- | --- | --- | --- |
| Before retraction | 169 (≥9) | 291 (≥2) | 1,003 (≥4) |
| After retraction | 144 (≥4) | 193 (≥3) | 793 (≥4) |

However, it would be inappropriate to directly compare all highly mentioned articles to all non-mentioned articles, as the two groups likely differ in key characteristics. For example, highly mentioned articles are more likely to be published in high-impact journals or be more recent publications. To address this, we applied coarsened exact matching (CEM) to construct a matched control group of non-mentioned retracted articles for each of the six highly mentioned article sets.

Although several matching methods can perform this task, such as propensity score matching (Rosenbaum & Rubin, 1983) and entropy balancing matching (Hainmueller, 2012), CEM was selected for its ability to control covariates to approximate or exact equality (Blackwell et al., 2009). CEM attempts to make treatment and control groups comparable by matching them for a set of covariates. It transforms continuous covariates into discrete categories and then identifies exact matches within these categories, helping to achieve a more precise balance (Iacus et al., 2012). However, due to the requirement for exact matching within the coarsened intervals, unmatched samples are discarded, potentially resulting in significant sample attrition, particularly when the number of covariates is large or when the coarsened intervals are overly fine-grained. To mitigate this issue, we selected the following four key covariates and controlled the coarsened intervals within reasonably narrow ranges:

- Subject field: The influence of retraction on articles may vary between fields. We controlled the subject fields of the treatment and control groups to be identical. We used the "Fields" classification in OpenAlex[1], which has 26 broad categories. This level of granularity seemed appropriate. Although more granular classifications are available (e.g., 252 subfields, 4,516 topics in OpenAlex), choosing these would make it difficult to match other covariates effectively.

---

[1] https://help.openalex.org/hc/en-us/articles/24736129405719-Topics

- Publication year: Publications from earlier periods may have undergone less stringent review processes or been disseminated through narrower channels, potentially leading to prolonged delays in the recognition of issues necessitating retraction. These delays could result in a retraction pattern that contrasts significantly with that of more recent articles (Van Noorden, 2011). We therefore ensured that the publication years of the treatment and control groups differed by no more than two years, following the matching methodology of Peng et al. (2022). While controlling for the same publication year would have been ideal, it would severely reduce the sample size given the need to balance other covariates, undermining the statistical robustness of the results.

- Journal citation rate: The level of attention varies between journals, with articles in higher-impact journals generally attract more media attention and citations after retraction (Nabavi, 2022). In this study, we controlled for journal impact using the rank percentile of the total citations-to-publications ratio of the publishing journal prior to the article's publication. Rank percentiles were preferred over absolute values because the latter varied widely, which could hinder effective matching. The difference in journal citation rate rank percentiles between the treatment and control groups was controlled to within 20%.

- Citations of most-cited author: An author's reputation can also influence the impact of a retraction (Trikalinos et al., 2008). Articles by authors with substantial reputations may take longer to be retracted and may continue to receive citations even after retraction. In this study, authors' reputation for an article was measured by the citation rank percentile of its most-cited author, determined by this author's position among the most-cited authors of all retracted articles. The citation data was obtained from OpenAlex in August 2024. The difference in rank percentiles of citation counts for the most-cited author between the treatment and control groups was kept within 20%, following the methodology of Peng et al. (2022).

After applying CEM, we assessed the balance of covariates between the control and treatment groups. We observed an improvement in balance, as evidenced by reductions in the L1 distances of covariates after matching (Appendix Tables A1 and A2).

The retraction time lag was included as a covariate in the *After retraction* part (Figure 1), as it might also influence the citations of articles (Sotudeh et al., 2022). The difference in retraction time lag between the treatment and control groups was controlled to within two years.

We used a one-to-three matching approach, where each highly mentioned retracted article was matched with three non-mentioned retracted articles both before and after retraction.

The descriptive statistics for the aforementioned variables are presented in Table 5. The average retraction time lag for articles in our dataset is approximately 2.5 years.

**Table 5**. Descriptive statistics of retracted articles (n = 15,461).

| Variable | Mean | Str. Err. | Min | Max |
|---|---|---|---|---|
| Publication year | 2017.754 | 0.023 | 2012 | 2021 |
| Retraction year | 2020.267 | 0.022 | 2012 | 2024 |
| Retraction time lag (days) | 917.903 | 6.238 | 1 | 4335 |
| Average journal citation rate | 13.726 | 0.122 | 0 | 268.159 |
| Most-cited author's total citations (in the year before article publication) | 864.927 | 29.294 | 0 | 64170 |

*2.4 Media attention and retraction time lags*

To examine the association between the level of news and social media attention and the retraction time lag, we initially applied the Shapiro-Wilk test to assess the normality of the retraction time lag data. We found that the data did not exhibit a normal distribution; therefore, we employed the Wilcoxon signed-rank test to evaluate the statistical significance of the differences in retraction time lags between highly and non-mentioned articles.

*2.5 Media attention and post-retraction citation changes*

To examine the association between the level of post-retraction news and social media attention and the effectiveness of retractions – as reflected by descreased citation rates – we constructed regression models for analysis.

To determine the appropriate regression model, we first tested for zero inflation in the citation data. The proportion of zero citations in news, blogs, and X was 39.9%, 44.1%, and 46.4%, respectively, indicating a significant degree of zero inflation that could compromise the assumptions of conventional count data models. Instead, a zero-inflated regression model was deemed more appropriate.

Since the dependent variable – citation counts – is a non-negative count variable, the two most commonly used regression models for such data are Poisson regression and negative binomial regression. To determine the better fit for this study, we conducted an overdispersion test on the citation count data (Table 6). The natural logarithm of the dispersion parameter ($\alpha$) was significantly greater than 1 in the zero-inflated negative binomial regression models for citations of articles with high and no mentions in news, blogs, and X, indicating that the citation data exhibit overdispersion. This suggests that the zero-inflated negative binomial (ZINB) regression model is more suitable than the zero-inflated Poisson (ZIP) regression model. We further evaluated goodness-of-fit metrics, which also indicated that the ZINB model showed a superior fit compared to the ZIP model (Appendix Table A3). Therefore, we selected ZINB regression models for citation analysis.

**Table 6**. Overdispersion test results for the citation data.

| Citation counts | $\alpha$ (s.e.) | p-value |
| --- | --- | --- |
| Articles with high and no news mentions | 2.244 (0.113) | 0.000 |
| Articles with high and no blog mentions | 2.461 (0.112) | 0.000 |
| Articles with high and no X mentions | 2.392 (0.066) | 0.000 |

ZINB regression models account for overdispersion and excess zeros in count data, providing more accurate estimates compared to standard models. ZINB regression models separate the data into two processes: one for generating zeros and the other for modeling count values, improving model fit and interpretation (Greene, 1994). In this study, excess zeros can be interpreted as "non-citable" aritcles – those that may not be cited due to being outdated or having a niche topic, rather than articles that simply have not yet been cited but might be in the future. One limitation of this approach is the potential for endogenous influences – articles with certain characteristics may be more likely to get cited, leading to biased estimates. To address this issue, we employed the CEM method to control for key characteristics of the treatment and control groups, thereby mitigating potential biases. This is necessarily approximate for citations due to publication delays, since a published citation may appear long after a retraction, even if added to an article long before the retraction.

The variables included in our regression models are listed in Table 7. The coefficient of the *Group × Post* interaction term is the main focus of this study, as it captures the difference in citation changes before and after retraction between highly mentioned and non-mentioned articles. To refine the model, we set *Post* as the zero-inflation term to help identify two mechanisms within the data: one representing the high probability of zero citations due to retraction, and the other involving the count component to explain normal citation variations. This allows for a more accurate analysis of the actual changes in citations after retraction.

**Table 7**. Variables used in the ZINB regression model to investigate factors influencing citation

counts.

| Variables | Explanation |
| --- | --- |
| Citation counts | Dependent variable. The number of citations the article received in a specific year. |
| Group × Post | Interaction term. Measures the difference in pre- and post-retraction citation changes between highly mentioned and non-mentioned articles. |
| Group | Binary variable indicating whether the article was highly mentioned by news outlets or social media. |
| Post | Binary variable indicating whether the article was retracted in a specific year. |
| Publication year | Year in which the retracted article was originally published. |
| Retraction time lag | Number of days between the article's publication and its retraction. |
| Journal citation rate | Ratio of citations to publications for the journal in which the article was published, measured prior to the article's publication. |
| Citations of most-cited author | Total citations received by the most-cited author of the article, measured one year prior to the article's publication. |

## 3. Results

*3.1 Do articles with high pre-retraction news and social media attention get retracted faster?*

Retracted articles with high pre-retraction attention from news outlets, blogs, or X were generally retracted more quickly than those without such attention. The difference was statistically significant for news and X, but not for blogs (Figure 2). This finding is intuitive, as articles that attract high levels of media attention are more likely to be scrutinized and discussed publicly, potentially leading to the faster identification of issues in the research (Haunschild & Bornmann, 2021; Zheng et al., 2024). News outlets have the potential to reach a wider audience, and X provides a faster and more direct platform for scientific discussions, allowing for swift detection and dissemination of concerns regarding problematic research. While the same may be true for blogs, their lack of a significant effect on retraction speed may primarily be due to smaller sample sizes instead of weaker or non-existent associations.

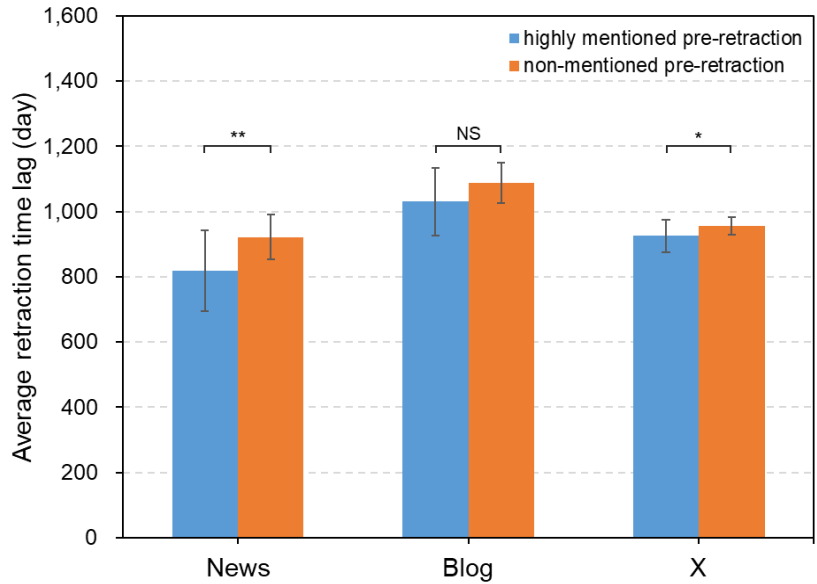

**Figure 2**. Mean retraction time lag (in days) for highly mentioned versus non-mentioned articles in news, blogs, or X. Note: *p = 0.05; **p = 0.01; ***p=0.001; the same below.

We examined several hypotheses to investigate why blog mentions might have little influence on retraction speed. To explore this, we analyzed the sentiment of mentions in news, blogs, and X using the SentiStrength software to determine whether they included negative commentary about the retracted articles. SentiStrength, which uses a dictionary-based approach, is designed to detect sentiment expressed in social media texts (Thelwall et al., 2010, 2012). It utilizes a dictionary of sentiment terms, emoticons, and additional rules to identify more complex expressions of sentiment, such as negation. We found that most mentions before retraction across news, blogs, and X were indeed negative, as classified by SentiStrength (Table 8), suggesting that they likely discussed problems with the articles, potentially speeding up the retraction process.

**Table 8**. Proportion of articles with negative mentions in news, blogs, and X before retraction.

| Source | Proportion of negative mentions |
| --- | --- |
| News | 151/169 (89.3%) |
| Blogs | 228/291 (78.4%) |
| X | 705/1,003 (70.3%) |

Additionally, we investigated the distribution of retraction reasons among retracted articles (Table 9). Our findings indicate that highly mentioned articles in news, blogs, and X were more frequently associated with mistakes (60.9%, 44.0%, and 37.7%, respectively) – a reason that tends to be linked to faster retractions, as reflected by shorter retraction time lags. Compared to non-mentioned articles, highly mentioned articles had a lower proportion of retraction reasons associated with longer retraction time lags, such as self-plagiarism, falsification and manipulation. The absence of a significant effect for blogs on retraction speed may be because blogs tend to focus more on misconduct-related retractions rather than retractions due to mistakes. Since misconduct-related retractions typically require longer investigative processes, involving author disputes and journal-led investigations, these retractions tend to be slower, as reflected in the mean retraction time lag column in Table 9.

**Table 9**. Proportion of retraction reasons for highly mentioned and non-mentioned retracted articles in news, blogs, or X. The total percentage exceeds 100%, as some articles were retracted for multiple reasons.

| | % of reasons of retraction | | | | | | |
| --- | --- | --- | --- | --- | --- | --- | --- |
| | News | | Blogs | | X | | |
| Retractrion reason | highly mentioned | non-mentioned | highly mentioned | non-mentioned | highly mentioned | non-mentioned | Mean retraction time lag (day) |
| Self-plagiarism | 4.1% | 17.9% | 8.2% | 21.2% | 14.2% | 22.9% | 1291.12 |
| Falsification and manipulation | 6.5% | 16.8% | 14.8% | 20.0% | 11.9% | 22.1% | 1141.68 |
| Reasons not available or uncategorizable | 22.5% | 28.6% | 25.1% | 29.9% | 28.4% | 27.2% | 1005.92 |
| Authorship issues | 1.8% | 4.1% | 1.7% | 4.9% | 3.0% | 5.9% | 903.82 |
| Ethical issues | 5.9% | 4.3% | 7.6% | 4.0% | 5.4% | 5.0% | 856.87 |
| Plagiarism | 0.6% | 5.1% | 5.2% | 6.9% | 6.4% | 6.8% | 820.00 |
| Mistakes | 60.9% | 32.0% | 44.0% | 23.8% | 37.7% | 24.8% | 783.13 |

*3.2 Do citations to articles with high news and social media attention decline more rapidly after retraction?*

Before analysing post-retraction citation trend, a parallel trends test was conducted to examine whether pre-retraction citation trajectories for highly mentioned and non-mentioned articles followed similar patterns. The results (Appendix Figure A1) indicate that pre-retraction citation trends did not significantly differ between highly mentioned and non-mentioned articles in news, blogs, and X, as confirmed by the regression model results (Appendix Table A4). This suggests that any post-retraction differences in citations can be reasonably associated with, if not attributed to, the level of news and social media attention.

The original citation counts of both highly mentioned or non-mentioned articles in news and social media followed similar trends: low annual citation counts prior to retraction, a sudden and significant increase in citations during the year of retraction, followed by a gradual year-by-year decline (Figure 3a). The abnormal citation rise in the retraction year has also been observed in Lu et al. (2013) and Sotudeh et al. (2022), and may be due to the "cited time lag" – the delay between adding a reference to an academic paper and its formal publication in a journal (Jiang et al., 2023).

Given the zero inflation and overdispersion in citation data, raw citation counts may not fully reflect actual citation patterns. Therefore, we applied the ZINB regression model for further analysis. The results of the regression models show that highly mentioned articles undergo an increase in average citation counts after retraction, whereas the average citation counts of non-mentioned articles do not demonstrate significant changes following retraction (Figure 3b). This pattern may be due to the greater visibility of highly mentioned articles, which attract more citations (Özkent, 2022). Researchers may cite these highly mentioned retracted articles after their retraction to engage in discussions or offer critiques (Tang, 2023).

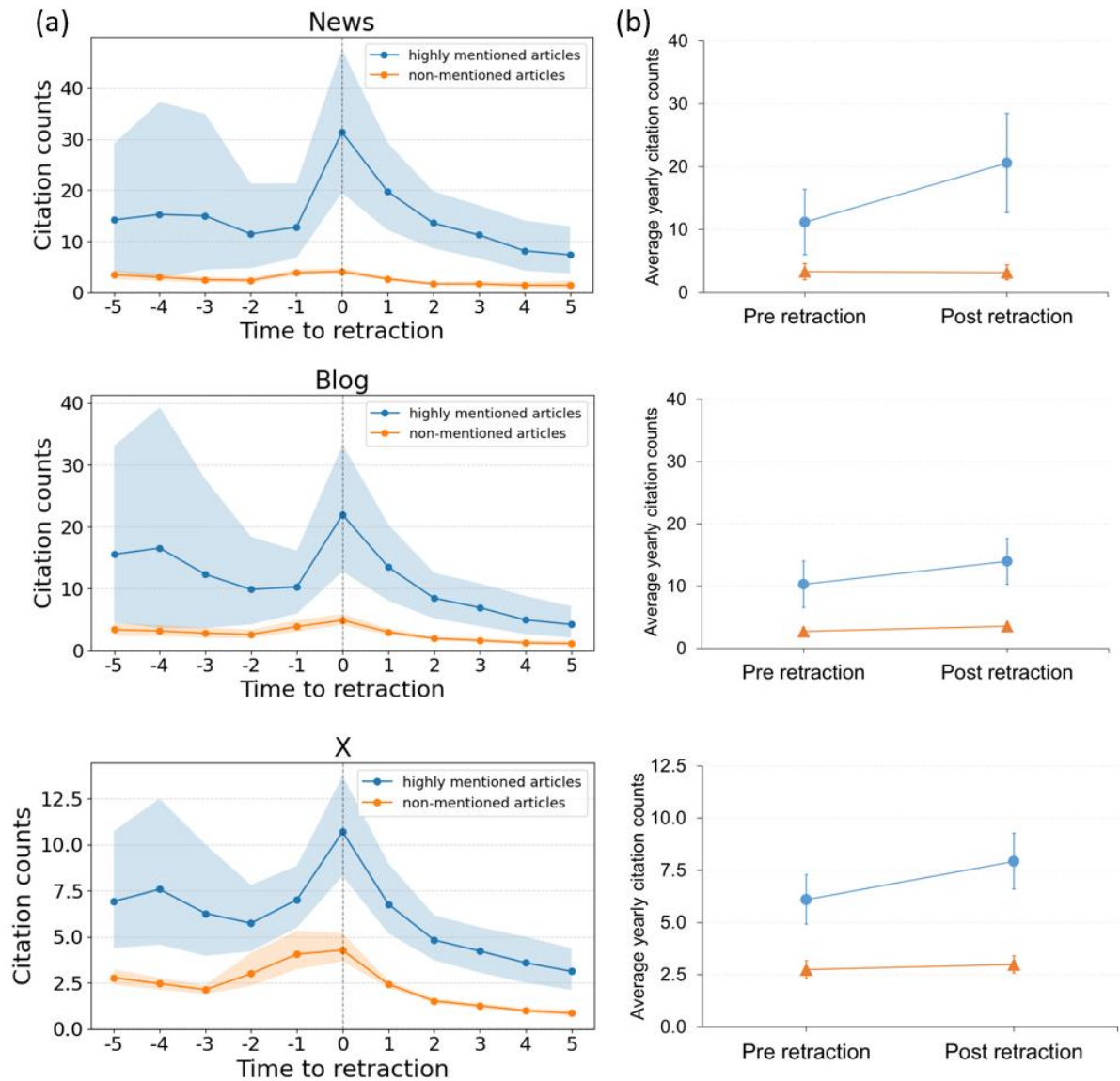

**Figure 3.** (a) Trends in citation counts; (b) Average yearly citation counts pre- and post-retraction for retracted articles with high or no news or social media attention as reflected by ZINB models. Citation counts in the retraction year are not included in either the pre-retraction or post-retraction periods. Error bars represent 95% confidence intervals.

For articles with frequent news mentions, the increase in post-retraction citations is significantly greater than those without mentions. In contrast, the citation changes between articles with high and no mentions on blogs and X do not have statistically significant differences before and after retraction (as indicated by the "group#post" term in Table 10). The zero-inflation term ("post") indicates a significant decrease in the likelihood of articles being "non-citable" after retraction for articles with high news, blog, and X mentions. Such articles are more likely to receive additional citations after retraction, potentially due to widespread discussions, criticism, or cautionary attention in both the media and academic communities.

**Table 10**. ZINB regression for citation counts for articles in the high attention and control sets.

|  | News | Blog | X |
|---|---|---|---|
| Independent variable | Coef. (S.E.) | Coef. (S.E.) | Coef. (S.E.) |
| group#post | 0.650 (0.190)*** | 0.048 (0.197) | 0.178 (0.093) |
| group | 1.211 (0.156)*** | 1.317 (0.169)*** | 0.797 (0.075)*** |
| post | -0.041 (0.103) | 0.259 (0.115)* | 0.084 (0.056) |
| pub_year | 0.022 (0.016) | 0.113 (0.016)*** | 0.076 (0.007)*** |
| retract_time_lag | 0.001 (0.000)*** | 0.001 (0.000)*** | 0.001 (0.000)*** |
| journal_cit_rate | 0.017 (0.002)*** | 0.017 (0.002)*** | 0.014 (0.001)*** |
| most_cited_author_cits | 0.000 (0.000)*** | 0.000 (0.000) | 0.000 (0.000)*** |
| cons | -45.576 (32.414) | -227.735 (32.023)*** | -152.9151 (15.082)*** |
| Zero-inflation |  |  |  |
| post | -21.635 (1.180)*** | -26.500 (0.797)*** | -25.799 (0.623)*** |
| cons | -3.444 (1.167)** | -2.998 (0.775)*** | -3.419 (0.622)*** |

It is perhaps surprising that articles with many news mentions experience a greater increase in citations after retraction compared to those without mentions. We found that most retraction notices of highly mentioned articles were covered in news, blogs, and X (Table 11). This suggests that exposure to retraction notices likely influence how these articles are cited post-retraction. Some researchers who encountered the retraction notices may have critically cited the retracted articles to discuss issues such as research ethics, methodological flaws, or disciplinary controversies, thereby serving as a warning to peers. Such citations also include those originating from editorials. Conversely, researchers who were not exposed to the retraction notices may have continued citing the retracted articles as if they remained valid, contributing to the observed increase in post-retraction citations, particularly for highly mentioned articles.

**Table 11**. Proportion of post-retraction mentions of retraction notices.

| Source | Proportion of mentioned retracton notices |
|---|---|
| News | 114/144 (79.2%) |
| Blogs | 185/193 (95.9%) |
| X | 713/793 (89.9%) |

## 4. Discussion

*4.1 News and social media have the potential to accelerate the retraction of flawed articles*

Social media can not only increase the awareness of scientific articles (Shamsi et al., 2022) and expand the impact of academic achievements (Ladeiras-Lopes et al., 2022), but they can also function as a monitoring mechanism that signals concerns of unreliable research (Peng et al., 2022; Prasad & Ioannidis, 2022). This study provides further empirical evidence supporting the notion that news and social media attention can expedite the retraction process for flawed articles. Based on these findings, we recommend that scientists use online platforms to raise concerns about problematic research in a timely manner.

While some X posts may spread misinformation regarding retracted articles (Abhari et al., 2023), they can also raise awareness of problems (Haunschild & Bornmann, 2021; Zheng et al., 2024). We found statistically significant evidence of this effect for news and X, but the results were inconclusive (in the same direction but not statistically significant) for blogs. A possible explanation for the effectiveness of news and X in accelerating retractions is their ability to rapidly disseminate information and amplify scientific discourse. This aligns with previous research showing that media coverage of hazards and risks can amplify risk perceptions (Kasperson et al., 1988; Sarathchandra & McCright, 2017). In the context of academic retractions, increased public scrutiny and discussion may pressure journals, editors, and institutions to act more quickly in investigating and retracting problematic research.

Our findings suggest that encouraging researchers to openly discuss potential issues in articles through social media and other platforms may help expedite the retraction process, enhance awareness of academic integrity issues, and safeguard the credibility of the scientific community.

*4.2 News and social media may not reduce the negative scholarly impact of retracted articles*

Retraction notices seem to receive less news coverage compared to the original articles (Barnett & Doblin, 2021; Serghiou et al., 2021). However, our analysis shows that most post-retraction mentions in news, blogs, and X about highly mentioned retracted articles referenced the retraction notices, thereby alluding to the retracted status of the articles. Despite this, our results show that highly mentioned articles experienced a greater post-retraction citation increase than non-mentioned articles. This effect was statistically significant for news mentions, while the direction was the same but not statistically significant for blog and X mentions. This increase is likely driven by a substantial number of critical citations discussing the retraction, reflecting on the underlying reasons for the retraction and the lessons learned from it. Also, it seems likely that only a tiny percentage of active researchers will see coverage of most retractions (De Cassai et al., 2022; Minetto et al., 2024). Consequently, some researchers may unknowingly continue citing these highly mentioned retracted articles, unaware of their retracted status. Although retractions should be clearly indicated on the publisher's website, researchers may refer to previously downloaded copies of articles rather than the live versions. Therefore, the overall effect of news and social media coverage on future citations is debatable.

Additionally, news and blogs covered only a small proportion of retracted articles in our study, which is consistent with earlier findings (Barnett & Doblin, 2021; Zeitoun & Rouquette, 2012; Zhang & Grieneisen, 2013). Scientific articles in general have very low coverage in news and blogs (Ortega, 2019; Peng et al., 2022). In the case of retracted articles, one reason for this might be that retraction issues conflict with the public perception of scientists as serious, reliable, and trustworthy figures. Moreover, reporting on scientific fraud and plagiarism is particularly challenging (Zeitoun & Rouquette, 2012). Therefore, the effectiveness of news and blogs in reducing the negative impact of retracted articles may be limited. In partial contrast, X covers a larger number of retracted articles after retraction, but typical X posts probably reach a smaller audience, which also limits its ability to warn potential citers about problematic research.

Potentially citing authors would presumably normally discover that an article has been retracted by seeing the retraction notice on the journal website above the article (or pasted onto it). Of course, this might be overlooked if the article had previously been downloaded or if it was accessed from a preprint repository or other source instead. The results suggest that, publicity does not seem to reach these authors so an alternative solution is needed. For example, publisher reference formatting systems and reference management software could be configured to run automatic checks for retractions and warn authors that have cited a retracted article. This is technically feasible with Retraction Watch data shared by Crossref, for example.

*4.3 Limitations*

There are several limitations of this study. First, 2% of retracted articles lacked corresponding retraction notices in the Retraction Watch database, preventing us from obtaining key details such as retraction dates, retraction time lags, retraction reasons, or post-retraction citations. This led to their exclusion from the analysis. Second, other factors may also influence the citation patterns of retracted articles, such as author collaboration (Zhang et al., 2020) and open access status (Zheng et al., 2024). However, including too many covariates in the matching process would have significantly reduced the sample size, potentially undermining the robustness of the study. Therefore, we selected the most relevant covariates based on previous studies (Peng et al., 2022). Third, this study did not distinguish between critical, neutral, and supportive citations of retracted articles. Citations that explicitly acknowledge retraction (e.g., discussing the reasons for retraction) play a valuable role in the scientific self-correction process. Future research would benefit from differentiating between negative citations that highlight retractions and citations that do not mention the retraction status, as this would provide a better understanding of how retracted articles continue to be cited.

## 5. Conclusions

The findings of this study suggest that news and social media attention can help mitigate the negative impact of retracted articles, primarily by reducing the retraction time lag. However, rather than reducing post-retraction citations, media coverage appears to be associated with their increase. One possible explanation is that media exposure enhances the visibility of retracted articles, making them more likely to be cited. These citations may arise from critical discussions explicitly addressing the retraction (e.g., in editorials and meta-analyses), but they may also include citations that fail to acknowledge the retracted status of the article. Although no cause-and-effect relationship has been shown, the potential role of news and social media in accelerating the retraction process suggests that increased public scrutiny can enhance the efficiency of scientific self-correction. However, the observed increase in post-retraction citations highlights the persistence of retracted articles in the scientific record. While some of these citations may serve a constructive role in discussions of research integrity and methodological flaws, others may reflect unintentional citations by researchers unaware of the retraction.

**Declaration of competing interests**

The authors have no conflicts of interest to declare.

**Acknowledgements**

Zhichao Fang is financially supported by the National Natural Science Foundation of China (No. 72304274). Hui-Zhen Fu is supported by the National Social Science Foundation of China (No. 22CTQ032). Er-Te Zheng is financially supported by the GTA scholarship from the University of Sheffield Information School. Mike Thelwall is supported by the Fundação Calouste Gulbenkian European Media and Information Fund (No. 187003). The authors thank Retraction Watch and Altmetric.com for providing the data for research purposes.

# Appendix

*Balance test of covariates before and after matching*

(1) Before retraction

**Appendix Table A1.** L1 distance of covariates between highly mentioned and non-mentioned articles before and after CEM (Pre-retraction). L1 distance values range between 0 and 1, with higher values indicating greater imbalance.

| Covariates | Media | Before matching | After matching |
|---|---|---|---|
| field | News | 0.38 | 0 |
|  | Blog | 0.28 | 0 |
|  | X | 0.27 | 0 |
| pub_year | News | 0.14 | 0.09 |
|  | Blog | 0.27 | 0.09 |
|  | X | 0.15 | 0.10 |
| journal_cit_rate | News | 0.42 | 0.17 |
|  | Blog | 0.34 | 0.13 |
|  | X | 0.32 | 0.14 |
| author_cit | News | 0.43 | 0.17 |
|  | Blog | 0.36 | 0.14 |
|  | X | 0.29 | 0.11 |

(2) After retraction

**Appendix Table A2.** L1 distance of covariates between highly mentioned and non-mentioned articles before and after CEM (Post-retraction).

| Covariates | Media | Before matching | After matching |
|---|---|---|---|
| field | News | 0.39 | 0 |
|  | Blog | 0.25 | 0 |
|  | X | 0.21 | 0 |
| pub_year | News | 0.19 | 0.06 |
|  | Blog | 0.36 | 0.12 |
|  | X | 0.26 | 0.10 |
| retract_year | News | 0.34 | 0.17 |
|  | Blog | 0.44 | 0.15 |
|  | X | 0.35 | 0.30 |
| journal_cit_rate | News | 0.36 | 0.13 |
|  | Blog | 0.34 | 0.15 |
|  | X | 0.28 | 0.20 |
| author_cit | News | 0.31 | 0.14 |
|  | Blog | 0.36 | 0.15 |
|  | X | 0.24 | 0.21 |

*Goodness-of-fit test*

We compared the goodness-of-fit metrics of the ZINB model and ZIP model (Appendix Table A3). The log-likelihood (LL) values were higher for the ZINB model than for the ZIP model, while the Akaike Information Criterion (AIC) and Bayesian Information Criterion (BIC) values were lower for the ZINB model than for the ZIP model, indicating that the ZINB model had a better fit. We therefore used the ZINB regression model for citation analysis.

**Appendix Table A3.** Goodness-of-fit test for ZINB and ZIP models

|  | LL | AIC | BIC |
|---|---|---|---|
| 1.News |  |  |  |
| ZINB | -7616.81 | 15255.64 | 15323.10 |
| ZIP | -21814.80 | 43649.59 | 43710.92 |
| 2.Blog |  |  |  |
| ZINB | -9934.16 | 19890.32 | 19961.55 |
| ZIP | -26522.85 | 53065.70 | 53130.46 |
| 3.X |  |  |  |
| ZINB | -36756.08 | 73534.16 | 73620.46 |
| ZIP | -75972.93 | 151965.90 | 152044.30 |

*Parallel trends test*

We set the value t = −1 (one year before retraction) as the baseline period to ensure that our comparisons focused on the immediate effects of retraction relative to the most recent pre-retraction period. The results indicate that before retraction, the citation changes between highly mentioned and non-mentioned articles in news, blogs, and X did not significantly differ. This confirms that the parallel trends assumption holds, suggesting that any post-retraction differences can reasonably be attributed to media attention levels.

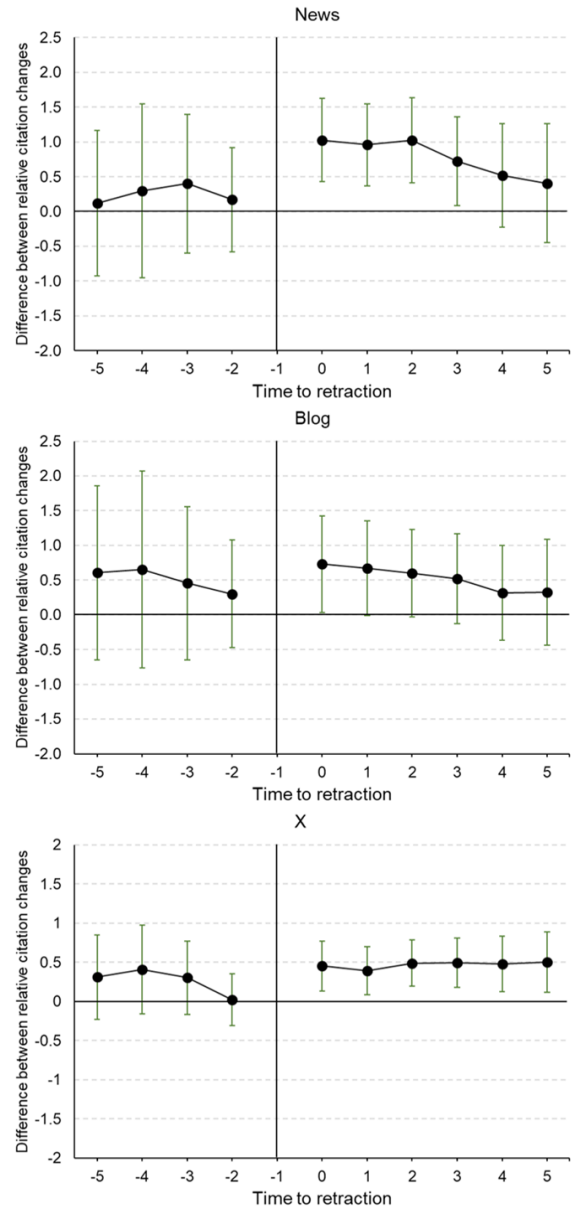

**Appendix Figure A1.** Parallel trends test for retracted articles highly and non-mentioned in news, blogs, and X

*Difference in citations between highly and non-mentioned retracted articles*

**Appendix Table A4.** Difference in citations between highly and non-mentioned retracted articles

|  | News | Blog | X |
|---|---|---|---|
|  | mean diff (s.e.) | mean diff (s.e.) | mean diff (s.e.) |
| Group#time(-5) | 0.122 (0.533) | 0.608 (0.640) | 0.311 (0.274) |
| Group#time(-4) | 0.299 (0.636) | 0.656 (0.724) | 0.407 (0.290) |
| Group#time(-3) | 0.404 (0.508) | 0.456 (0.563) | 0.303 (0.238) |
| Group#time(-2) | 0.171 (0.381) | 0.300 (0.395) | 0.023 (0.168) |
| Group#time(0) | 1.026 (0.305) *** | 0.729 (0.356) * | 0.453 (0.164) ** |
| Group#time(1) | 0.958 (0.301) *** | 0.672 (0.349) | 0.393 (0.158) * |
| Group#time(2) | 1.024 (0.310) *** | 0.600 (0.322) | 0.491 (0.151) *** |
| Group#time(3) | 0.723 (0.324) * | 0.519 (0.329) | 0.495 (0.160) ** |
| Group#time(4) | 0.519 (0.380) | 0.318 (0.348) | 0.480 (0.180) ** |
| Group#time(5) | 0.407 (0.436) | 0.328 (0.390) | 0.504 (0.196) ** |